\documentclass[12pt]{article}
\pdfoutput=1
\usepackage{natbib}
\usepackage{multirow}
\usepackage{amsthm}
\usepackage{amssymb}
\usepackage{xcolor}
\usepackage{bbm}
\usepackage{lscape}
\usepackage{caption}
\usepackage{subcaption}
\usepackage{mathtools}
\usepackage{url}
\usepackage{float}
\usepackage{amsmath}
\usepackage{mathalfa}

\usepackage{algorithm}
\usepackage[parfill]{parskip}
\usepackage{algorithmic}
\usepackage[T1]{fontenc}
\usepackage{titling}
\usepackage[toc,page]{appendix}
\usepackage{listings}
\usepackage{geometry}
\usepackage{dsfont}
\usepackage{wrapfig}
\usepackage{graphicx}
\usepackage[english]{babel}
\newtheorem{definition}{Definition}[section]
\newtheorem{theorem}{Theorem}[section]

\numberwithin{equation}{section}

\definecolor{ao}{rgb}{0.0, 0.5, 0.0}

\title{Design of platform trials with a change in the control treatment arm}
\author{Peter Greenstreet$^{1,2,\star}$, Thomas Jaki$^{3,4}$, Alun Bedding$^{5}$, Pavel Mozgunov$^{3}$
\\  \footnotesize{$^1$ Department of Mathematics and
Statistics, Lancaster University, Lancaster, UK} 
\\  \footnotesize{$^2$ Exeter CTU, University of
Exeter, Exeter, UK} 
\\  \footnotesize{$^3$ MRC Biostatistics Unit, University of
Cambridge, Cambridge, UK} 
\\  \footnotesize{$^4$ University of Regensburg, Regensburg, Germany} 
\\  \footnotesize{$^5$ Roche Products Ltd., Welwyn Garden City, UK}
\\  \footnotesize{$^\star$ p.greenstreet@lancaster.ac.uk} }
\date{}
\begin{document}
\maketitle

\begin{abstract}
Platform trials are a more efficient way of testing multiple treatments compared to running separate trials. In this paper we consider platform trials where, if a treatment is found to be superior to the control, it will become the new standard of care (and the control in the platform). The remaining treatments are then tested against this new control. In such a setting, one can either keep the information on both the new standard of care and the other active treatments before the control is changed or one could discard this information when testing for benefit of the remaining treatments. We will show analytically and numerically that retaining the information collected before the change in control can be detrimental to the power of the study. Specifically, we consider the overall power, the probability that the active treatment with the greatest treatment effect is found during the trial. We also consider the conditional power of the active treatments, the probability a given treatment can be found superior against the current control. We prove when, in a multi-arm multi-stage trial where no arms are added, retaining the information is detrimental to both overall and conditional power of the remaining treatments. This loss of power is studied for a motivating example.  We then discuss the effect on platform trials in which arms are added later. On the basis of these observations we discuss different aspects to consider when deciding whether to run a continuous platform trial or whether one may be better running a new trial.
\end{abstract}

%




%
\section{Introduction}
\label{sec:introduction}
Clinical trials take many years and are very costly to run \citep{MullardAsher2018Hmdp} which has therefore lead to multiple developments in methodology on how to efficiently design them \citep{PallmannPhilip2018Adic}. One of these developments has been the idea of platform trials in which multiple treatments are tested against a common control group \citep{UrachS.2016Mgsd, RoystonPatrick2003Ndfm, WasonJamesM.S.2012Odom, MaxineBennett2020Dfaa}. Platform trials can be advantageous due to having a shared trial infrastructure and shared control groups \citep{BurnettThomas2020Aeta}. The interest in these types of trials has increased since the beginning of the COVID-19 pandemic \citep{LeeKimMay2021Scwa, StallardNigel2020EADf}, as platform trials can result in therapies being identified faster while reducing cost and time~\citep{cohen2015adding}. 
\par
One additional ability one may want from a platform trial's design is to be able to change the control group to a beneficial new treatment found within the trial. A change of control has happened in multiple platform trial such as STAMPEDE  \citep{SydesMatthewR2012Ftdi} and RECOVERY \citep{HorbyPeter2021DiHP}. When changing the control group one may think of using all the data collected to calculate the future test statistics. There is little work currently investigating whether keeping the data collected prior to the change of the control group treatment is the most efficient approach. In this paper we will consider two settings: (i) We keep all the concurrent data from before the change in control; (ii) We do not keep any of the data prior to the control changing. Concurrent data means only participants recruited to the current control arm at the same time as the active arm of interest are used in the comparisons. For work on non-concurrent controls see recent work by \cite{lee2020including, MarschnerIanC2022Aoap, SavilleBenjaminR2022TBTM, WangChenguang2022ABmw}.
\par
This work will focus on multi-arm multi-stage  trials (MAMS) in which additional treatments can be planned to be added at multiple points during the trial. Multi-arm trials allow for multiple treatments to be compared at once against a common control treatment. Multi-stage trials have interim analyses which allow for ineffective treatments to be dropped for futility (or lack of benefit) earlier. As a result interim analyses can improve a trials operating characteristics \citep{PocockStuartJ.1977GSMi, ToddSusan2001Iaas}. They also can allow treatments to stop early if a superior treatment is found, however in the case studied here the first time this happens this superior treatment will become the new control. 
\par
We will focus our investigation on two types of power: 
(i) Conditional power of a treatment - the probability a given treatment can be found superior against the current control. (ii) Overall power of the trial - the probability that the active treatment with the greatest treatment effect is found during the trial. 
\par
In Section \ref{Sec:Method} we introduce the notation, the null hypotheses of interest and  discuss type I error. Section \ref{Sec:Cp} studies the conditional power for a general design, where treatments can be added at different points in a preplanned manner. Then Section \ref{Sec:Cp} gives theorems to when keeping the old data is guaranteed to be detrimental in MAMS trials where all the treatments begin at the same time. In Section \ref{Sec:OP} we give the formulation for the overall power along with its definition and give theorems to when keeping the old data is guaranteed to be detrimental. A motivating example is then studied in Section \ref{Sec:Motivating} for both the case when arms begin the trial at the same time and also when one starts later. Finally we discuss the considerations one needs to make when deciding whether to use the pre-change data or not.

\section{Notation and type I error control}
\label{Sec:Method}
Consider a clinical trial with up to $K$ experimental arms that will be tested against one common control arm. The primary outcome on each patient is independent and normally distributed with known variance $\sigma^2$. Each active treatment is tested at $J$ analyses with $J-1$ interim analyses. Let $n_{k,j}$ denote the number of patients recruited to treatment $k$ by the end of its $j^\text{th}$ stage assuming that recruitment of this arm had begun at the start. For this paper the focus will be on equal sample size and allocation ratio for each treatments and equally spaced interim analyses for all treatments, as this ensures equal pairwise error for each treatment without needing to have multiple boundary shapes \citep{GreenstreetPeter2021Ammp}. 
Therefore the number of patients recruited between interim analyses is equal i.e. $n_{k,j} - n_{k,j^\star}=n_{k^\star,j} - n_{k^\star,j^\star}$ for all $k, k^\star = 0, \hdots, K$ and $j, j^\star = 0, \hdots, J$. Let $n_{k,0}$ define the number of patients already recruited to an active treatment that started the trial before treatment $k$ enters the trial.  We have $k_{n_{k',j'}}'$ denoting the current control treatment at point $n_{k',j'}$, where $j'$ is the stage for treatment $k'$ where it became the control, with $k' = 0, \hdots, K$ and $j' = 0, \hdots, J$. Therefore $n_{k',j'}$ denotes the number of patients recruited prior to treatment $k'$ becoming control at its $j'$ \hspace{-2mm} $^\text{th}$ stage. For simplicity we drop the subscript from $k_{n_{k',j'}}'$ as the focus of this work will be on only changing the control group once, with, $k'=0$ at the beginning of the trial. 
\par
The null hypotheses of interest are
$
H_{k'1}: \mu_1 \leq \mu_{k'}, H_{k'2}: \mu_2 \leq \mu_{k'}, ... , H_{k'K}: \mu_K \leq \mu_{k'},
$
where $\mu_1, \hdots, \mu_K$ are the mean responses on the $K$ experimental treatments and $\mu_{k'}$ is the mean response of the current control group with $\mu_{0}$ being the mean response of the initial control. Each of the $K$ hypotheses is potentially tested at a series of analyses indexed by $j=\ddot{j}_{k,k',j'}+1,\hdots,J$ where $\ddot{j}_{k,k',j'}$ is the last stage for $k$ before $k'$ became the control. When one is in the case that all the treatments begin at once then $\ddot{j}_{k,k',j'}=j'$ as each interim for each treatment happens at the same time. However if treatments are added at different points this may not be the case. For example in a 3 arm trial if treatment 1 becomes the control at its first stage and treatment 2 is not added till after treatment 1's first analysis then $\ddot{j}_{2,1,1}=0$. At analysis $j$ for treatment $k$, to test $H_{k'k}$ it is assumed that responses, $X_{k,i}$ and $X_{k',i'}$, from patients $i=n_{k,0},\hdots, n_{k,j}$ and $i'= n_{k',0},\hdots, n_{k,j}$ are observed respectively. These hypotheses are tested at given analysis $j$ using the test statistic:
\begin{align*}
Z_{k,k',j}=& \frac{ \sum^{n_{k,j}}_{i=\max(n_{k,0},n_{k',0})+1} X_{k,i} - \sum^{n_{k,j}}_{i=\max(n_{k,0},n_{k',0})+1} X_{k',i} }{\sigma\sqrt{2(n_{k,j}-\max(n_{k,0},n_{k',0}))}}.
\end{align*}
In order to ensure only concurrent controls are used  we have $\max(n_{k,0},n_{k',0})$.
If only the data post the change in the control is used the test statistics are:
\begin{equation*}
Z^\star_{k,k',j,j'}= \frac{ \sum^{n_{k,j}}_{i=\max(n_{k,0},n_{k',0},n_{k',j'})+1} X_{k,i} - \sum^{n_{k,j}}_{i=\max(n_{k,0},n_{k',0},n_{k',j'})+1} X_{k',i} }{\sigma\sqrt{2(n_{k,j}-\max(n_{k,0},n_{k',0},n_{k',j'}))}},
\end{equation*}
\par
where $\max(n_{k,0},n_{k',0},n_{k',j'})$ includes the point in which the control changes and only if $n_{k,0},n_{k',0} \leq n_{k',j'}$ will $Z^\star_{k,k',j,j'} \neq Z_{k,k',j}$.
These test statistics are used to test $H_{k'k}$. Upper and lower stopping boundaries,  $U=(u_{1},\hdots,u_{J})$ and $L=(l_{1},\hdots,l_{J})$, are used for the decision-making as follows. If $ Z_{k,k',j}> u_{j}$ then $H_{k'k}$ is rejected and the conclusion that treatment $k$ is superior to the current control is made. If $ Z_{k,k',j}< l_{j}$ then treatment $k$ is dropped from all subsequent stages of the trial. If the $Z$ statistics for all the treatments fall below their lower boundary, the trial stops for futility. Treatment $k$ and control continues to its next stage if $l_{j} \leq Z_{k,k',j} \leq u_{j}$. If the post change data is only used the same rules apply now replacing $Z_{k,k',j}$ with $Z^\star_{k,k',j,j'}$.
If multiple treatments exceed their given upper boundary at the same time point, for example treatments $k_1$ and $k_2$ where $k_1, k_2 = 1,\hdots K$,  then one finds $Z_{k_1,k_2,j}$ and if this is negative then one takes forward treatment $k_2$ as the new control treatment and otherwise $k_1$ is taken. 
\par
These upper and lower stopping boundaries are group-sequential bounds which are pre-defined in order to control the original type I error control aimed for in the original trial. Therefore for example they could be aiming to control the pairwise error rate (PWER) \citep{WasonJamesMS2014Cfmi, Choodari-OskooeiBabak2020Anea}, the family wise error rate (FWER) \citep{BurnettThomas2020Aeta, MagirrD.2012AgDt, GreenstreetPeter2021Ammp} or the false discovery rate (FDR) \citep{RobertsonDavidS2022Oecf, CuiXinping2023SPfP}. Typically when continuing to use the same boundary as already pre-defined there is no longer a guarantee this will control the type I error of interest after the change. 
This is because the original bounds were not designed for this. 
\par

\section{Conditional power}
\label{Sec:Cp}
\begin{figure}[]
  \centering
  \includegraphics[width=0.6\linewidth,trim= 0 1.5cm 0 1.5cm, clip]{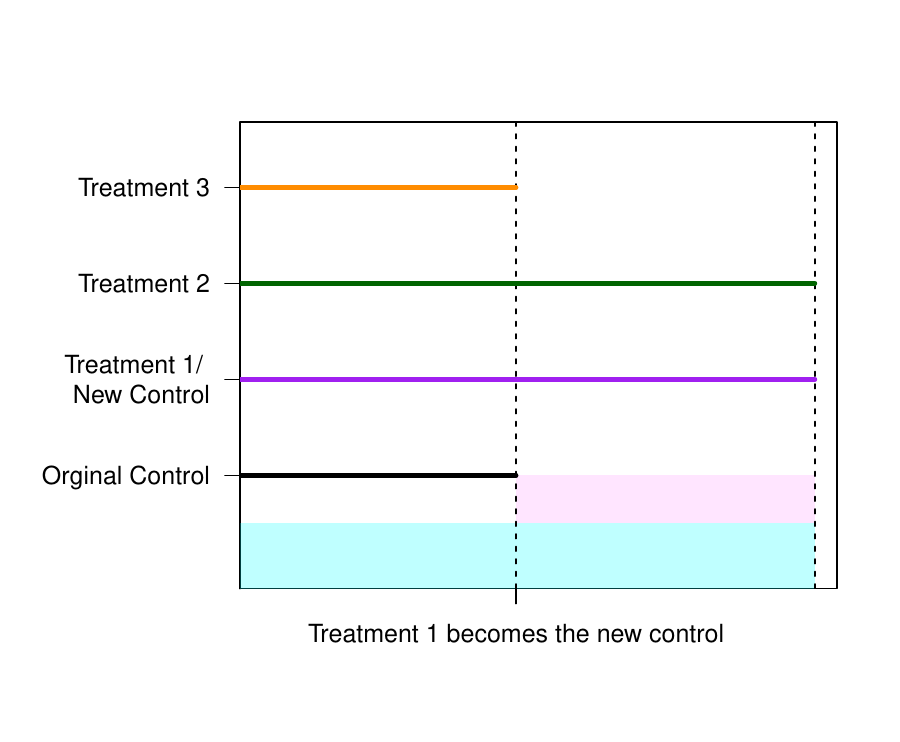} 
  \caption{Illustration of the difference in area of data used for conditional power when comparing using all the data to only the data post change. The area highlighted in blue represents the data used if all the data is retained. The area in pink represents the data if only the data post the change in control is used. }
  \label{fig:Cexample} 
\end{figure}

The conditional power for a given treatment $k^\star$ is the probability that given treatment $k'$ is the new standard of care after its $j'^{\text{th}}$ stage that treatment $k^\star$ is found superior to the new control $k'$, when tested for treatment $k^\star$ remaining analyses. The conditional part of conditional power can be split into 3 events. Event 1 ($E^1_{k^\star,k',j'}$) is the event that treatment $k'$ becomes the control at its $j'^{\text{th}}$ stage. Event 2 ($E^2_{k^\star,k',j'}$) is that treatment $k^\star$ is still in the trial when treatment $k'$ becomes the control. Event 3 ($E^3_{k^\star,k',j'}$) is that none of the other $k$ treatments become the control. The detailed formulations for $E^1_{k^\star,k',j'}$, $E^2_{k^\star,k',j'}$, $E^3_{k^\star,k',j'}$ are given in the Appendix \ref{App:Events}. The conditional power is therefore defined as:
\begin{definition}
The conditional power for treatment $k^*$ for given $k'$ and $j'$ is 
\begin{equation*}
P(\text{reject } H_{k'k^\star}| E^1_{k^\star,k',j'} \cap E^2_{k^\star,k',j'} \cap E^3_{k^\star,k',j'}).
\end{equation*}
\label{Def:Conpower}
\end{definition}
From Definition \ref{Def:Conpower} the conditional power is the probability we reject $H_{k'k^\star}$ given that at $j'$ stage for treatment $k'$ it became the control and treatment $k^\star$ is still being tested.
%
%
%
Figure \ref{fig:Cexample} shows the difference in the data included in the calculation of the conditional power based on the motivating example discussed in Section \ref{Sec:Motivating}. The motivating example has 2 stages and 4 arms to begin the trial. In this figure it is assumed that treatment 3 has stopped for futility after the first stage and treatment 1 has become the new control at this stage. Therefore the conditional power of interest is that treatment 2 is found superior to treatment 1, the new control. The area highlighted in blue represents the data used if all the data is retained. This therefore covers the whole length of the trial. Whereas the area in pink is if only the data post the change in control is used therefore only covers the second stage of the trial. 
\par
\par
In order to equate the conditional power one can use the conditional probability definition to remove the need to calculate any highly truncated normal distributions. The conditional power is:

\begin{align*}
\begin{cases}
0 & \text{ if } n_{k^\star,J} \leq n_{k',j'} \\ 
\frac{P(E^1_{k^\star,k',j'} \cap E^2_{k^\star,k',j'} \cap E^3_{k^\star,k',j'} \cap E^4_{k^\star,k',j'})}{P(E^1_{k^\star,k',j'} \cap E^2_{k^\star,k',j'} \cap E^3_{k^\star,k',j'})} & \text{ if } n_{k^\star,J} > n_{k',j'}
\end{cases}. 
\end{align*}
where $E^4_{k',k^\star,j'}$ is the event that we $\text{reject } H_{k'k^\star}$ within the rest of the trial. The formulations for $E^1_{k^\star,k',j'}$, $E^2_{k^\star,k',j'}$, $E^3_{k^\star,k',j'}$ and $E^4_{k^\star,k',j'}$ are given in the Appendix \ref{App:Events}. This can be calculated using multivariate normal distributions as discussed for the motivating example in Section \ref{Sec:Motivating}, using the mean of each test statistic $Z_{k,k',j}$,
\begin{align*}
\frac{ (\mu_k -\mu_{k'})\sqrt{(n_{k,j}-\max(n_{k,0},n_{k',0}))} }{\sigma\sqrt{2}},
\end{align*}
and the correlation matrix, $\Sigma$. The correlation matrix can be split into multiple values, $\psi_{(k_1,k'_1,j_1),(k_2,k'_2,j_2)}$, that depend on the correlation between $Z_{k_1,k'_1,j_1}$ and $Z_{k_2,k'_2,j_2}$, and $\psi_{(k_1,k'_1,j_1),(k_2,k'_2,j_2)}$ equals,
\begin{equation*}
\begin{cases}
0 & \text{for } k_1 \neq k_2, k'_2  \; \& \; k'_1 \neq k_2, k'_2 
\\
\frac{\max(0,n_{k_1,j_1}-\max(n_{k_1,0},n_{k'_1,0},n_{k'_2,0}) )}{2\sqrt{(n_{k_1,j_1}-\max(n_{k_1,0},n_{k'_1,0}))(n_{k_1,j_2}-\max(n_{k_1,0},n_{k'_2,0})) }} 
& \text{for } k_1 = k_2  \; \& \; k'_1 \neq k'_2 \; \& \; n_{k_1,j_1} \leq n_{k_2,j_2}
\\
-\frac{\max(0,n_{k_1,j_1}-\max(n_{k_1,0},n_{k_2,0}, n_{k'_1,0}) )}{2\sqrt{(n_{k_1,j_1}-\max(n_{k_1,0},n_{k'_1,0}))(n_{k_2,j_2}-\max(n_{k_2,0},n_{k_1,0})) }} & \text{for } k_1 = k'_2  \; \& \; k'_1 \neq k_2 \; \& \; n_{k_1,j_1} \leq n_{k_2,j_2}
\\
\frac{\max(0,n_{k_1,j_1}-\max(n_{k_1,0},n_{k_2,0},n_{k'_1,0}) )}{2\sqrt{(n_{k_1,j_1}-\max(n_{k_1,0},n_{k'_1,0}))(n_{k_2,j_2}-\max(n_{k_2,0},n_{k'_1,0})) }} & \text{for } k_1 \neq k_2  \; \& \; k'_1 = k'_2  \; \& \; n_{k_1,j_1} \leq n_{k_2,j_2}
\\
\sqrt{\frac{n_{k_1,j_1}-\max(n_{k_1,0},n_{k'_1,0})}{n_{k_1,j_2}-\max(n_{k_1,0},n_{k'_1,0})}} & \text{for } k_1 = k_2  \; \& \; k'_1 = k'_2  \; \& \; n_{k_1,j_1} \leq n_{k_2,j_2}.
\end{cases} 
\end{equation*} 
It is worth noting that one does not need to consider the case that $k_2=k'_1$ when $n_{k_1,j_1}<n_{k_2,j_2}$ as it is not possible for a treatment to go from being a control back to being an active treatment. 
\par
In the case of only considering the data post changing the control, the test statistics before the change are now independent of the test statistics post the change. Therefore one only needs the 
event that we $\text{reject } H_{k'k}$ within the rest of the trial. For the case where only the post change data is used we define this as $E^{\star 4}_{k^\star,k',j'}$. If one is in the case where treatment $k^\star$ joins the trial after treatment $k'$ becomes the control then $E^{4}_{k^\star,k',j'}=E^{\star 4}_{k^\star,k',j'}$ as there is no data pre the change that is shared. The formulations for $E^{\star 4}_{k^\star,k',j'}$ is given in the Appendix \ref{App:Events}.  The conditional power in this case is:
\begin{align}
\begin{cases}
0 & \text{ if } n_{k,J} \leq n_{k',j'} \\ 
P(E^{\star 4}_{k^\star,k',j'}) & \text{ if } n_{k,J} > n_{k',j'}
\end{cases}.
\end{align} 
Once again this can be calculated using multivariate normal distributions as discussed for the motivating example in Section \ref{Sec:Motivating} using the mean of each test statistic $Z_{k,k',j}$,

\begin{align*}
\frac{ (\mu_k -\mu_{k'})\sqrt{(n_{k,j}-\max(n_{k,0},n_{k',0},n_{k',j'}))} }{\sigma\sqrt{2}};
\end{align*}
and the correlation matrix which can be split into multiple $\psi_{i,i^\star}$ that depend on the correlation between  $Z^\star_{k,k',j_1,j'}$ and $Z^\star_{k,k',j_2,j'}$, and equals, 
\begin{equation*}
\psi_{i,i^\star}= \begin{cases}
\sqrt{\frac{n_{k,j_1}-\max(n_{k,0},n_{k',j'}) }{n_{k,j_2}-\max(n_{k,0},n_{k',j'})}} &\text{for } j_1 \leq j_2.
\end{cases}
\end{equation*}
When one is in the case that all the treatments begin at once this simplifies the equations as now $n_{k,0}=0$ for all $k=0,\hdots K$. Additionally as shown below one can now prove when it is guaranteed that there is no benefit to retaining the information pre change in control treatment when considering conditional power and using the predefined boundaries. These can not be proven for when treatments are added later however as shown in Section \ref{Sec:Motivating} there can still be a negative effect from keeping the data pre the change in the control treatment. 

\subsection{When all the treatment start at the beginning of the trial}
When all treatments begin at the same time, it can be proven that for many cases there can never be benefit to retaining the information pre change in control treatment when considering conditional power and using the predefined boundaries. The first theorem (Theorem \ref{ConTheorem1}) states that if there is only one stage left and the upper boundary is positive, then keeping the historic data is detrimental to the conditional power.   
\begin{theorem}
If a treatment $k'$ becomes the control group treatment at stage $J-1$ ($E^1_{k^\star,k',J-1} \cap E^3_{k',k',J-1}$) and $u_J \geq 0$ then the conditional power for treatment $k^\star$ when retaining the data before the control changed is less than or equal to the conditional power for treatment $k^\star$ when not retaining the pre-change data.
%
\label{ConTheorem1}
\end{theorem}

The proof of Theorem \ref{ConTheorem1} is given in Appendix \ref{App:ConTheoremproves}. This theorem can be further extended. First in Theorem \ref{ConTheorem2} which states that if there are multiple stages of the trial left and both the upper and lower boundaries are greater than or equal to 0 then retaining the post change data is detrimental to the conditional power. The second extension is Theorem \ref{ConTheorem3} which states that if there are multiple stages of the trial left and the upper boundaries are positive and there is no lower boundaries then retaining the post change data is detrimental to the conditional power.

\begin{theorem}
If a treatment $k'$ becomes the new control group treatment at stage $j'-1$ ($E^1_{k^\star,k',j'-1} \cap E^3_{k',k',j'-1}$) and $u_j \geq 0$ and $l_j \geq 0$ for all $j=(j'+1) \hdots J$ then the conditional power for treatment $k^\star$ when retaining the data before the control changed is less than or equal to the conditional power for treatment $k^\star$ when not retaining the pre-change data. 
\label{ConTheorem2}
\end{theorem}

\begin{theorem}
If a treatment $k'$ becomes the new control group treatment at stage $j'-1$ ($E^1_{k^\star,k',j'-1} \cap E^3_{k',k',j'-1}$) and $u_j \geq 0$ and there are no lower boundaries for all $j=(j'+1) \hdots J$ then the conditional power for treatment $k^\star$ when retaining the data before the control changed is less than or equal to the conditional power for treatment $k^\star$ when not retaining the pre-change data.
\label{ConTheorem3}
\end{theorem}

The proof for Theorem \ref{ConTheorem2} is given in the Appendix \ref{App:ConTheoremproves}. The proof for Theorem \ref{ConTheorem3}, which is similar to the proof of Theorem \ref{ConTheorem2}, is given in the  Supplementary Materials Section 1. Furthermore as we will show in the Supplementary Materials Section 6 even if $l_j < 0$ for some $j=(j'+1) \hdots J$ then one will find that retaining the old information is likely detrimental for the conditional power. However in Supplementary Materials Section 7 it is shown that there are cases when $l_j < 0$ where keeping the old data can be beneficial for conditional power. 
\section{Overall power} 
\label{Sec:OP}
Overall power of a treatment gives the probability that during the trial the active treatment with the greatest positive treatment effect is either taken forward as the new control or is declared superior compared to a new control, if the control has already changed. Therefore overall power can be thought of as two main parts: either the correct treatment becomes the new control first, or another treatment becomes the new control and subsequently this treatment is found to be better than this new control. This gives the overall power definition as: 

\begin{definition}
The overall power for the treatment $k^\star$  which has the greatest treatment effect,  $\mu_{k^\star} \geq \mu_k \forall k= 1,\hdots, K,$ equals
\begin{equation*}
P( \bigcup^{J}_{j^\star=1}[ E^1_{k^\star,k^\star,j^\star} \cap E^3_{k^\star,k^\star,j^\star}] \cup \bigcup_{k'\in \{1,\hdots,K \}/k^\star } \bigcup^{J}_{j'=1}[  E^1_{k',k^\star,j'} \cap E^2_{k',k^\star,j'} \cap E^3_{k',k^\star,j'} \cap E^4_{k',k^\star,j'}] ).
\end{equation*}
\label{Def:Overpower}
\end{definition}


Due to the multiple disjoint sets within Definition \ref{Def:Overpower}, the overall power can be split into multiple, easy to compute, parts.   The first of these is the probability that at each interim $j^\star$,  treatment $k^\star$ becomes the control ($\Xi_{k^\star,j^\star} $) and this equals: 
\begin{equation}
\Xi_{k^\star,j^\star}=P(E^1_{k^\star,k^\star,j^\star} \cap E^3_{k^\star,k^\star,j^\star}).
\label{Equ:OP1}
\end{equation}
The probability another treatment becomes the new control and then this treatment is found to be better then the new control ($\Omega_{k^\star,k',j'}$) can be split into every possible $k'$ and $j'$. 
\begin{equation}
\Omega_{k^\star,k',j'}=\begin{cases}
0 & \text{ if } n_{k,J} \leq n_{k',j'} \\ 
P(E^1_{k',k^\star,j'} \cap E^2_{k',k^\star,j'} \cap E^3_{k',k^\star,j'} \cap E^4_{k',k^\star,j'}) & \text{ if } n_{k,J} > n_{k',j'}
\end{cases}.
\label{Equ:OP2}
\end{equation}
Combining Equation (\ref{Equ:OP1}) and Equation (\ref{Equ:OP2}) the overall power is: 
\begin{equation*}
\sum^{J}_{j^\star=1} \Xi_{k^\star,j^\star} + \sum_{k'\in \{1,\hdots,K \}/k^\star }\sum^{J}_{j'=1}\Omega_{k',k^\star,j'}.
\end{equation*}
When we consider only using the data post change in control the probability another treatment becomes the new control and then this treatment is found to be better then the new control ($\Omega^\star_{k^\star,k',j'}$) becomes:
\begin{equation*}
\Omega^\star_{k^\star,k',j'}=\begin{cases}
0 & \text{ if } n_{k,J} \leq n_{k',j'} \\ 
P(E^1_{k^\star,k',j'} \cap E^2_{k^\star,k',j'} \cap E^3_{k^\star,k',j'})P(E^{\star 4}_{k^\star,k',j'}) & \text{ if } n_{k,J} < n_{k',j'}
\end{cases}
.
\end{equation*}
 This is due to the independence of event 4 with the rest of the events.
Therefore the overall power is:
\begin{equation*}
\sum^{J}_{j^\star=1} \Xi_{k^\star,j^\star} + \sum_{k'\in \{1,\hdots,K \}/k^\star }\sum^{J}_{j'=1}\Omega^\star_{k^\star,k',j'}.
\end{equation*}
As shown below one can now prove when it is guaranteed that there is no benefit to retaining the information pre change in control treatment when considering overall power and using the predefined boundaries. These can not be proven for when treatments are added later, however as shown in Section \ref{Sec:Motivating} there can still be a negative effect from keeping the data pre the change in the control treatment. 
\subsection{When all the treatment start at the beginning of the trial}
In the scenario where all the treatments begin at once, from Theorem \ref{ConTheorem2} and Theorem \ref{ConTheorem3} for the conditional power, one can prove similar results for the overall power. 
\begin{theorem}
If $u_j \geq 0$ and $l_j \geq 0$ for all $j=1, \hdots, J$ then the overall power when retaining the data before the control changed is less than or equal to the overall power when not retaining the pre-change data.
%
%
\label{OverTheorem1}
\end{theorem}
\begin{theorem}
If $u_j \geq 0$ and there are no lower boundaries for all $j=1, \hdots, J$ then the overall power when retaining the data before the control changed is less than or equal to the overall power when not retaining the pre-change data.
\label{OverTheorem2}
\end{theorem}
The proof for Theorem \ref{OverTheorem1} and Theorem \ref{OverTheorem2} is given in Appendix \ref{App:OverTheoremproves}.
Furthermore as is shown in Supplementary Materials Section 6 even if $l_j < 0$ for any $j=1, \hdots, J$ then there are cases that retaining  information pre the change in the control group  is detrimental for the overall power. This is shown in the example in  the Supplementary Materials Section 6 as the difference in conditional power between keeping and discarding the pre change data is negative, therefore, so will the overall power. 
\section{Motivating trial example}
\label{Sec:Motivating}
We consider the motivating trial of TAILoR \citep{PushpakomSudeep2020TTaI}. The TAILoR trial was a 4 arm trial which studied the effect of different doses of a treatment on HIV. The study had 1 interim analysis. We are going to use the operating characteristics from this study to see the effects on overall and conditional power if the control was changed mid trial if a treatment was found superior. In the original design the family wise error rate (FWER) \citep{PushpakomSudeepP2015TaIR} was controlled at 5\% one sided for a normal continuous endpoint and there was a planned 90\% power. The trial was planned to have equal allocation across stages. In addition the clinically relevant effect of $\theta_1= 0.545$ and uninteresting effect $\theta_0= 0.178$ assuming the variance $\sigma^2=1$ was used.
\par
Triangular stopping boundaries will be used \citep{WhiteheadJ.1997TDaA} as recommended in \cite{WasonJamesM.S.2012Odom}. The stopping boundaries when all the treatments start at once will be calculated using the approach given in \cite{MagirrD.2012AgDt} to control FWER for the design before the change in control. In addition we will consider the design if one of the treatments was added at the end of the first stage. Therefore the stopping boundaries will be found using the approach given in \cite{GreenstreetPeter2021Ammp} to control FWER for the design before the change in control. The calculations of the power will be done using \cite{greenstreet2023preplanned} in-order to control the pairwise power for each treatment. This is chosen as it is similar to that used in the original trial but is designed for trials which continues after a treatment is taken forward.  
The calculations were carried out using R \citep{Rref} with the method given here having the multivariate normal probabilities being calculated using the package \texttt{mvtnorm} \citep{mvtnorm}; the upper and lower boundaries when all the treatments start at once where found using \texttt{MAMS} \citep{jaki2019r} and the code was parallelised using packages \texttt{doParallel} \citep{doParallel} and \texttt{foreach} \citep{foreach}. The code is available at \textit{https://github.com/pgreenstreet/change\_control\_platform}.
\par
\begin{figure}[H]
\centering
\begin{subfigure}{0.49\textwidth}
  \centering
  \includegraphics[width=1\linewidth,trim= 0 0.5cm 0 2cm, clip]{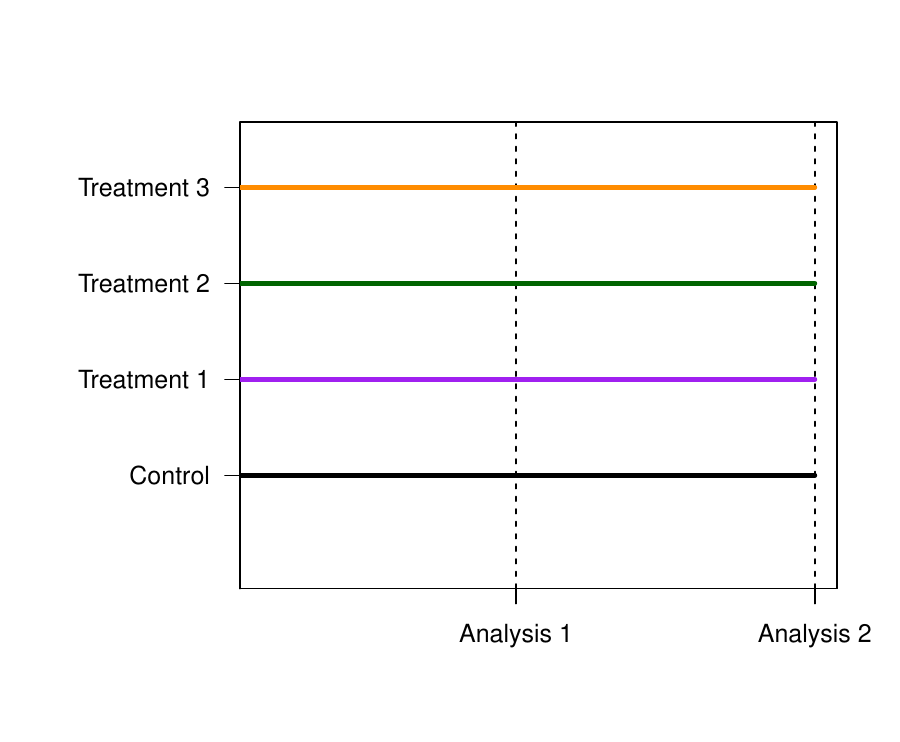}  
\caption{}
              \label{fig:Case1}
\end{subfigure}
\begin{subfigure}{0.49\textwidth}
  \centering
  \includegraphics[width=1\linewidth,trim= 0 0.5cm 0 2cm, clip]{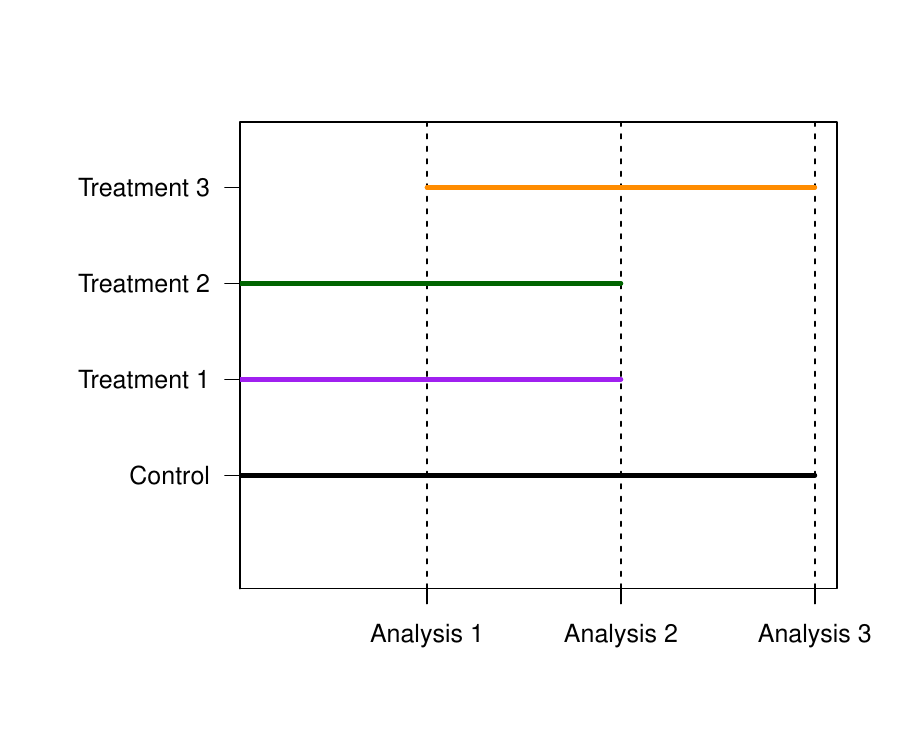}  
  \caption{}
              \label{fig:Case2}
\end{subfigure}
\caption{Illustration of the motivating trial. Figure \ref{fig:Case1} illustrates when all treatments start at the beginning and Figure \ref{fig:Case2} illustrates when one treatment starts at the end of the first stage.}
              \label{fig:Case1Case2}
\end{figure}

This section will be split into two parts: the first case (Case 1) will look at the case when all treatment start at the beginning; the second (Case 2) will look at the case where one of the treatments is added a stage later. Case 1 and Case 2 are depicted in Figure \ref{fig:Case1} and Figure \ref{fig:Case2} respectively.
\par

\subsection{Case 1: All treatments start at the same time}
\label{subsec:Case1}
Using the approach by \cite{MagirrD.2012AgDt} the triangular upper and lower stopping boundaries are found to be
\begin{equation*}
U= \begin{pmatrix}
2.330 & 2.197\\
2.330 & 2.197\\
2.330 & 2.197
\end{pmatrix}, \; \; \; L= \begin{pmatrix}
0.777 & 2.197\\
0.777 & 2.197\\
0.777 & 2.197
\end{pmatrix}. 
\end{equation*}
Using \cite{greenstreet2023preplanned} the maximum sample size is 344 based on 43 patients per arm per stage to ensure pairwise power of 90\%. Due to each treatment getting the same number of treatments per stage we define $n=n_{k,1}$ for all $k =0,1,2,3$ therefore the number of patients recruited at the second stage for a treatment which runs for both stages is $n+n=2n$. This subsection begins by discussing the formulation of the equations for both conditional power and overall power. After this we then study the results of the conditional power and overall power under different treatment effects and compare the effect of using all available data compared to only using the data post change.
\subsubsection{Conditional power}
\label{Case1:CP}
There is only one place in the trial where the conditional power is not zero as there is only one interim analysis in the study and all the treatments begin at the same point. This happens when a treatment becomes the new control at the first stage. We define the treatment of interest as $k^\star$, define the new control as $k'$ and define the other treatment in the study as $k_1$. The conditional power for treatment $k^\star$ when treatment $k'$ becomes the new control at stage 1 is:
\begin{equation}
\frac{P(E^1_{k^\star,k',1} \cap E^2_{k^\star,k',1} \cap E^3_{k^\star,k',1} \cap E^4_{k^\star,k',1})}{P(E^1_{k^\star,k',1} \cap E^2_{k^\star,k',1} \cap E^3_{k^\star,k',1})}.
\label{eq:conMAMS}
\end{equation}
When calculating Equation (\ref{eq:conMAMS}) one can take advantage of the fact that all the treatment start at the same time. Therefore, $Z_{k,k',j'}<0 \cup Z_{k,0,j'}<u_{j'}$ can be simplified to $Z_{k,k',j'}<0$. This is because testing $Z_{k,0,j'}>u_{j'}$ and  $Z_{k,k',j'}<0$ is equivalent to testing $Z_{k,k',j'}=Z_{k,0,j'}-Z_{k',0,j'}<0$  and $Z_{k',0,j'}>u_{j'}$ for treatment $k'$ to be taken forward when all the treatments start at the same point.
When we only retain the new information the conditional power is
\begin{equation}
P(E^{\star 4}_{k^\star,k',1}). 
\label{equ:CPnew}
\end{equation}
In the Supplementary Materials Section 2 the formulations used to calculate Equation (\ref{equ:CPnew}) and Equation (\ref{eq:conMAMS}) are given.
\subsubsection{Overall power}
\label{Case1:OP}
To calculate overall power in addition to the calculations above one needs the probability the treatment of interest $k^\star$ becomes the control at stage 1 or stage 2 of the trial. The two other arms in this case are defined as $k_1$ and $k_2$. Due to all the arms starting at the same point this simplifies the calculation of both $\Xi_{k,1}$ and $\Xi_{k,2}$. The complete formulation for $\Xi_{k,1}$ and $\Xi_{k,2}$ is given in the Supplementary Materials Section 3. Using the calculations for the conditional power one can find both $\Omega_{k^\star,k',1}$ and $\Omega^\star_{k^\star,k',1}$. The overall power for treatment $k^\star$ when the old information is retained is
\begin{equation*}
\sum^{2}_{j^\star=1} \Xi_{k^\star,j^\star} + \sum_{k'\in \{1,2,3\}/k^\star }\Omega_{k^\star,k',1}.
\end{equation*}
When only new data is used the overall power is
\begin{equation*}
\sum^{2}_{j^\star=1} \Xi_{k^\star,j^\star} + \sum_{k'\in \{1,2,3\}/k^\star }\Omega^\star_{k^\star,k',1}.
\end{equation*}
\subsubsection{Results}

\begin{figure}[]
\centering
              \includegraphics[width=0.6\linewidth]{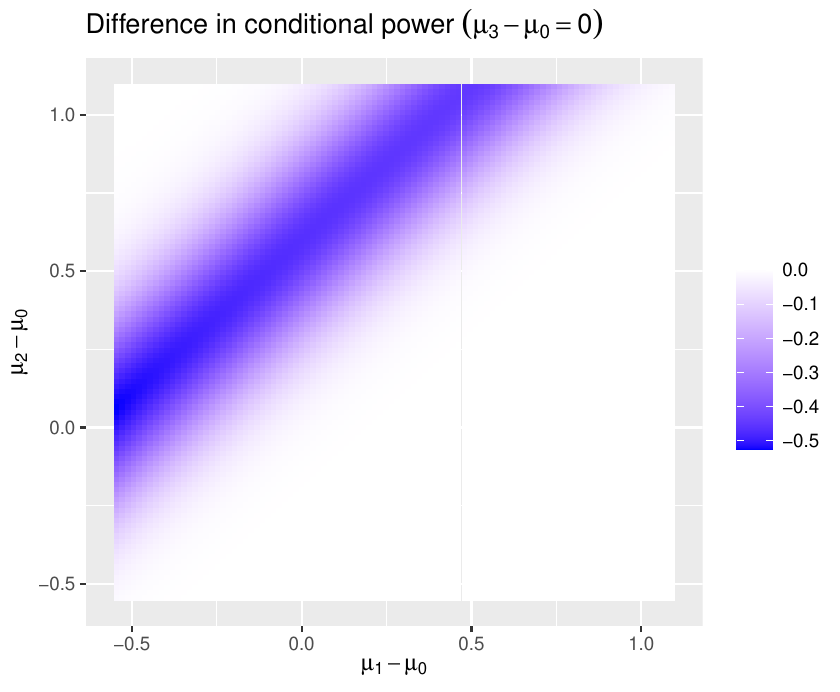}
              \caption{The difference in conditional power between keeping the data pre change and not, for treatment 2 given that treatment 1 has gone forward at the first stage.}
              \label{fig:CpDMp}
\end{figure}

In Figure \ref{fig:CpDMp} the difference between conditional power when retaining all the old data and not retaining the data can be seen. The conditional power for treatment 2 when treatment 1 is the new control after the first stage is studied. The y-axis gives the treatment effect of treatment 2 compared to the original control treatment. The x-axis gives effect of treatment 1 compared to the original control treatment. The colour as given on the scale, to the right of the figure, defines the difference in conditional power between retaining the information pre the change and not. The effect of different values of $\mu_3$ is very small. As it has very little effect on the probability that treatment 2 is found superior to treatment 1 in the final stage. Therefore we will focus on the results for when $\mu_3-\mu_0=0$. However in the Supplementary Materials the effect of $\mu_3-\mu_0$ having an uninteresting treatment effect is shown.
\par 
As shown in the Supplementary Materials Section 4, for the conditional power when all the data is retained, if the difference between  $\mu_2$ and $\mu_1$ is greater than 1, the probability that treatment 2 is found superior to the treatment 1 is at least 88.9\%. When $\mu_2$ is less then $\mu_1$ the probability of incorrectly taking treatment 2 forward is at worst 0.01\%.  This is in comparison to when only the data post the change is retained. Now when the difference between $\mu_2$ and $\mu_1$ is greater than 0.760 we have conditional power of above 90\%. When $\mu_2$ is less than $\mu_1$ the probability of incorrectly taking treatment 2 forward is at worst 1.15\%. The figures illustrating the conditional power for these can be seen in the Supplementary Materials Section 4.

\par    
The difference between conditional power when retaining all the old data and not retaining the data can be seen in Figure \ref{fig:CpDMp} when $\mu_3-\mu_0=0$. As can be seen in Figure \ref{fig:CpDMp} when $\mu_2-\mu_1$ is around 0.5 then the loss in conditional power is maximised. This can be greater than 50\%. However as this difference becomes a lot more extreme the loss becomes close to 0. 
This is because at this point either approach has almost an 100\% chance of finding treatment 2 superior to treatment 1. In the Supplementary Materials Section 5 we study the effect on conditional power of different possible values of $Z_{(1,0),1}$ and $Z_{(2,0),1}$ for  one of these points, $\mu_1=-0.25$ and $\mu_2= 0.75$. Here we can ignore the value of $Z_{(3,0),1}$ as this does not influence the probabilities as shown in the proof to Theorem 1 in Appendix \ref{App:ConTheoremproves}.  It is shown here that even in this case where there is on average very little benefit in only retaining the new information there are potential values of $Z_{(1,0),1}$  and $Z_{(2,0),1}$ where there is large benefit in only using the new data. However the probability of these $Z$ values happening is very small for the given $\mu_1$ and $\mu_2$. When $\mu_2-\mu_1 <0$ the difference in conditional power is small. This is because for both approaches the probability that treatment 2 is found superior to treatment 1 when in fact it is not is small. 
\par 
\par
The overall power is very similar between retaining the data or not as shown in Figure \ref{fig:OpD}. Once again the focus being $\mu_3-\mu_0$ equal to zero. However in the Supplementary Materials Section 4 the results are also shown when $\mu_3-\mu_0=0.178$ and $\mu_3-\mu_0=-0.178$. This is along with the figures illustrating the overall power for when the data is retained or not. The y-axis gives the treatment effect of treatment 2 compared to the original control and the x-axis gives effect of treatment 1 compared to the original control. The colour as given on the scale to the right defines the difference in overall power at that given point.
\par
\begin{figure}[]
          \centering
              \includegraphics[width=0.6\linewidth]{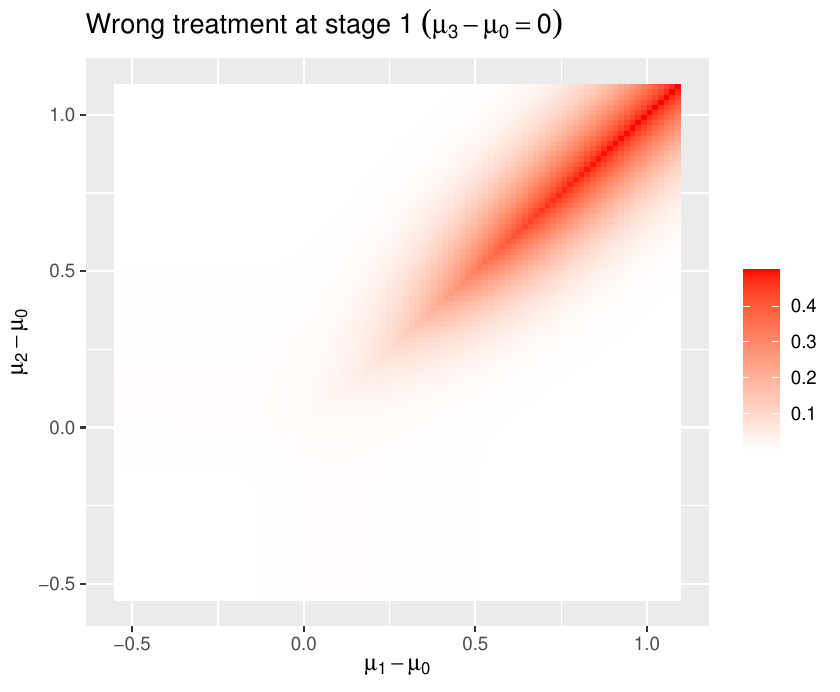}
              \caption{The probability of the treatment which does not have the greatest treatment effect becoming the control at the first stage.}
              \label{fig:OpW}
          \end{figure}

The maximum difference in overall power is 1.7\%. This is compared to a maximum change of 52.6\% for the difference in conditional power. This is because when calculating the overall power most of the time the correct treatment will be taken forward compared to the original control instead of one of the other treatments, however, when studying the conditional power this is not considered. Therefore when calculating the overall power the probability of a mistake is taken into account. This effect can be seen in Figure \ref{fig:OpW}. This figure gives the probability of the treatment which does not have the greatest treatment effect becoming the control at the first stage. This shows that in many of the areas, where the difference in conditional power was at its greatest, it is unlikely that treatment 1 or 3 would have been taken forward instead of treatment 2.

          \begin{figure}[]
          \centering
              \includegraphics[width=0.6\linewidth]{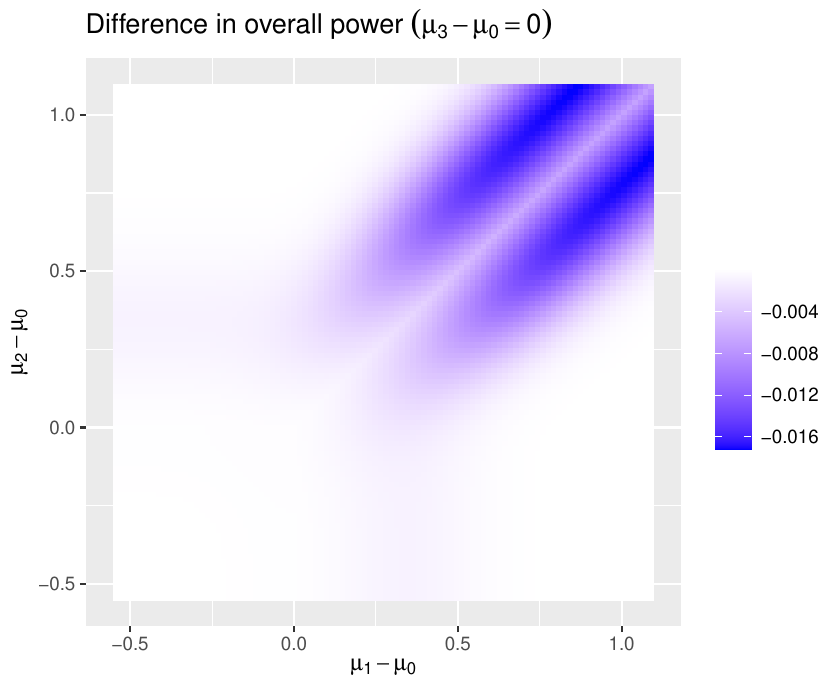}
              \caption{The difference in overall power between keeping the data pre change and not.}
              \label{fig:OpD}
          \end{figure}
     
\subsection{Case 2: A treatment added later}
\label{Subsec:Case2}
This subsection will study the case where one of the treatments is added a stage later as shown in Figure \ref{fig:Case2}. Using the approach by \cite{GreenstreetPeter2021Ammp} the triangular stopping boundaries are found to be
\begin{equation*}
U= \begin{pmatrix}
2.358 & 2.223\\
2.358 & 2.223\\
2.358 & 2.223
\end{pmatrix}, \; \; \; L= \begin{pmatrix}
0.786 & 2.223\\
0.786 & 2.223\\
0.786 & 2.223
\end{pmatrix}. 
\end{equation*}
Based on 43 patients per arm per stage the maximum sample size is now 387 in order to control the pairwise power at 90\% \citep{greenstreet2023preplanned}. This addition accounts for the patients which would need to be added for the later treatment as seen in Figure \ref{fig:Case2}. As each treatment gets the same number of treatments per stage we define $n=n_{k,1}-n_{k,0}=n_{k,2}-n_{k,1}$ for all $k =0,1,2,3$.
\par
We begin this subsection by discussing the equations for both conditional power and overall power. For both of these we will split the calculations into two. The first is for the treatments who begin the trial. The second is for the treatment which joins after 1 stage.
For brevity only an explanation of the calculation required is given below, however, in the Supplementary Materials Section 9 the equations required are given. After this we then study the results of the conditional power and overall power under different treatment effects and compare the effect of using all available data compared to only using the data post change.
\subsubsection{Conditional power}
\label{Conditional power Case 2}
The only non zero conditional power for the 2 treatments that start the trial are if a treatment becomes the control at the first stage. Therefore the conditional power when old data is retained is very similar to the one given in Subsection \ref{Case1:CP}, however now the 3rd treatment is no longer considered, as it is unable to become the control at stage 1 as it is yet to be tested. When considering the conditional power when only new data is retained then Equation (\ref{equ:CPnew}) is used as once again the conditional power is independent of the other continuing treatments.
\par
The conditional power for treatment 3 is non zero in two cases. The first is when the new control has been declared at the first stage of the trial - so at the point treatment 3 begins. In this case the conditional power is the same if the old data is retained or not. This is because only concurrent controls are used therefore there is no difference between the two as no old data is available. One can therefore take advantage of the fact that event $E^4_{k^\star,k',1}$ is independent of $E^1_{k^\star,k',1}$, $E^2_{k^\star,k',1}$ and $E^3_{k^\star,k',1}$. The conditional power given that treatment 1 or 2 becomes the control at its second stage is more numerically complex. However one can still use the fact that treatment 1 and 2 start at the same time therefore, $Z_{k_1,k',j'}<0 \cup Z_{k_1,0,j'}<u_{j'}$ can be simplified to $Z_{k_1,k',j'}<0$  in this case. The complete equations for these can be seen in the Supplementary Materials Section 9.  

\subsubsection{Overall power}
Calculating  $\Xi_{k,1}$ and $\Xi_{k,2}$ for the treatments that start at the beginning of the trial is very similar to the calculations given in Subsection \ref{Case1:OP}. However for $\Xi_{k,1}$ one only needs to consider the other treatment that began the trial at the beginning. Similar for $\Xi_{k,2}$ one only needs to consider the first stage for treatment 3. Therefore the overall power for treatment $k^\star$ given it is treatment 1 or 2 is
\begin{equation*}
\sum^{2}_{j^\star=1} \Xi_{k^\star,j^\star} + \Omega_{k^\star,k'=\{1,2\}/k^\star ,1}.
\end{equation*}
\par
If the third treatment is the superior treatment then one needs to calculate $\Xi_{k,1}$ and $\Xi_{k,2}$  accounting for the fact this treatment has been added at a later stage. One needs to include, all the possible outcomes for the other treatments that ensures that treatment 3 is taken forward first. As a result this requires 9 integrals and 4 integrals for $\Xi_{k,1}$ and $\Xi_{k,2}$, respectively. The reason for the reduction in integrals for the second is if treatment 3 becomes the new control at it second stage then this guaranties that the other treatments have gone below their lower boundaries at some point, where as this is not the case for $\Xi_{k,1}$. This can be seen clearly in the equations given in the Supplementary Materials Section 9. The overall power for treatment $k^\star$ given it is treatment 3 is 
\begin{equation*}
\sum^{2}_{j^\star=1} \Xi_{k^\star,j^\star} + \sum_{k' \in \{1,2\} } \sum^{2}_{j'=1} \Omega_{k^\star,k',j'}.
\end{equation*}
\subsubsection{Results}
\begin{figure}[]
\centering
\begin{subfigure}{0.49\textwidth}
  \centering
  \includegraphics[width=1\linewidth,trim= 0 0cm 0 0cm, clip]{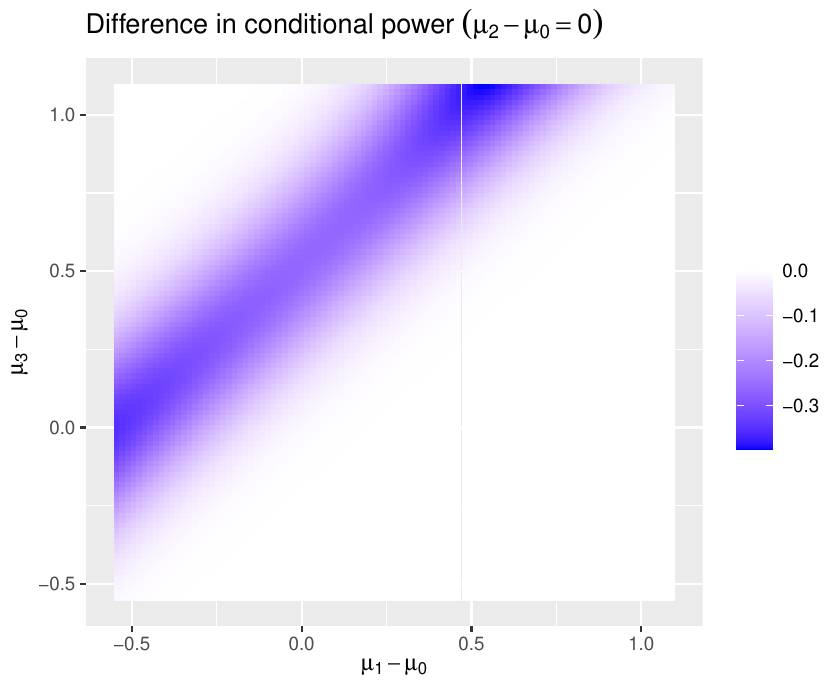}  
\caption{}
              \label{fig:CpDlate}
\end{subfigure}
\begin{subfigure}{0.49\textwidth}
  \centering
  \includegraphics[width=1\linewidth,trim= 0 0cm 0 0cm, clip]{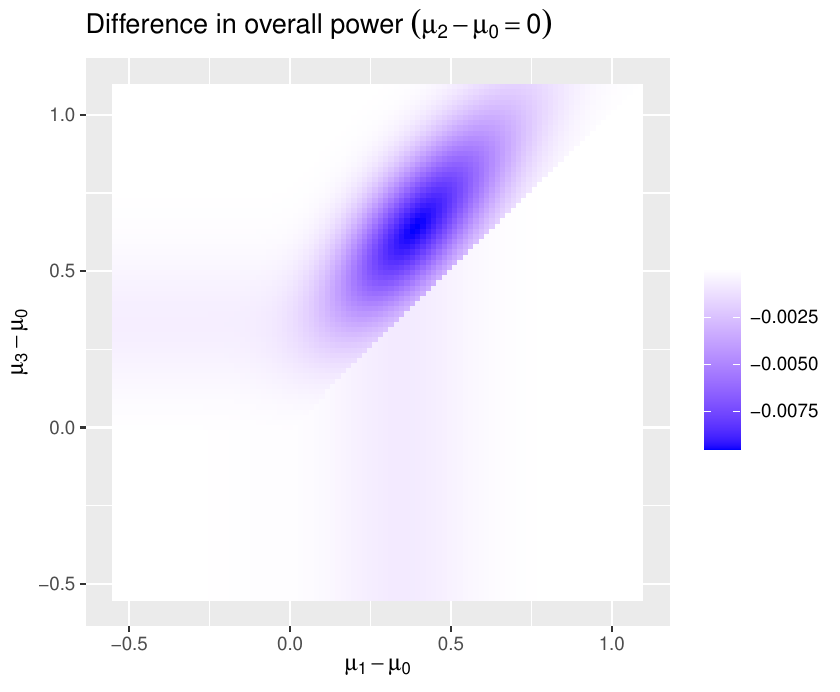}  
  \caption{}
              \label{fig:OpDlate}
\end{subfigure}
\caption{The difference in conditional power for treatment 3 given that treatment 1 has gone forward at the second stage (Figure \ref{fig:CpDlate}) and the overall power (Figure \ref{fig:OpDlate}) when treatment 3 starts after the first stage.}
              \label{fig:Dlate}
\end{figure}

In this section the main focus will be on the conditional and overall power of treatment 3. This is because as shown in the Supplementary Materials Section 10 the results for the conditional and overall power for the earlier treatments are almost identical to the case when all the treatments start at the same time as seen in Section \ref{subsec:Case1}. For the conditional power we are going to therefore focus on the case that treatment 1 becomes the new control at its second stage and treatment 2 has treatment effect equal to that of the original control. This is the focus as if treatment 1 becomes the control at its first stage there is no difference between the conditional power for treatment 3 if old data is retained or not. This is as the same data will be used in both cases, as only concurrent data is used. The difference in conditional power can be seen in Figure \ref{fig:Dlate}. 
\par
As can be seen here once again there is no benefit found in keeping the historic data. 
This is because, as detailed in the proof to Theorem \ref{ConTheorem1}, it is still unlikely that Equation (\ref{Eq:hatZabove}) is true given that treatment $k'$ became the control.  
However it is worth noting that the difference in power is less than it was for the case when all the treatments started at once with the maximum difference being 39.8\%.

\par
When we consider the difference in overall power we once again see that keeping the old data is detrimental as shown in Figure \ref{fig:Dlate}. However as seen in the case when all treatments start at the same time the effect of keeping the old data is a lot less for the overall power compared to the conditional power. Figure \ref{fig:Dlate} is no longer symmetric as the treatment effect of treatment 3 has no effect on the difference for treatment 1. This is because at the first opportunity treatment 3 could be taken forward, treatment 1 will be at its final analyses as discussed in Section \ref{Conditional power Case 2}.  
\par
Through this section it has been shown that for an example, even for the later arm, there is no benefit in keeping the data. However this is not always true. For example when considering the case when there is only one analysis for each treatment there is more likely to be benefit from keeping the old data for the later treatment as can be seen in the Supplementary Materials Section 11. 
\section{Discussion}
\label{Sec:Discussion}
In this paper we have studied the effect of keeping or discarding the data post a change in control in a platform study, with the focus being on the power of the study. This work has shown that in many cases one is likely to be better off not retaining the data, when using the same stopping boundaries. Therefore in this case it would be potentially beneficial to start a new trial. This would also give time for decisions with regards to which treatments should be compared to the new control. There are likely more benefits in starting a new trial including being able to adjust the research question, the target population, and the treatment dose as well as many more. However these would involve creating a new protocol and setting up a new trial which may be more administrative and logistical work compared to continuing the trial.  
\par
In Section \ref{Sec:Cp} and Section \ref{Sec:OP} it was shown for many cases when all arm start at the same time one can prove that the overall and conditional power will be lower by retaining the old data. However even in scenarios where one can not prove that the power will be lost by retaining the old data it has been shown in the Supplementary Materials Section 6 that one is still likely better of not retaining the previous data. This section looks at the effect of using the symmetric boundary shape of O'Brien and Fleming \citep{OBrienPeterC.1979AMTP} for a 3 stage example using the same operating characteristics as given in the motivating example of TAILor \citep{PushpakomSudeepP2015TaIR}. However in the Supplementary Materials Section 7 there is an example when all arm start at ones where keeping the old data can be beneficial. In Section \ref{Sec:Motivating} it was shown that the loss in conditional power can be very large when old data is retained. It was shown that for overall power the loss is less than that for conditional power, but there is still a loss in overall power.
\par
This work has shown that when adding additional treatments later that depending on the boundary shapes there may not be benefit in keeping the historic data for the later treatments as well as the ones which start the trial. However in this case it is not ensured that keeping the old data will be detrimental as is shown in the Supplementary Materials Section 11 for an example where each treatment gets one analysis. 
\par
\par
Equal allocation ratio between arms has been used in this work. This has been done for simplicity, so the pairwise error is equal for every treatment before the change in control, one could consider extending this work to consider different allocation ratios for each treatment and one could also extend this work to consider changing the control and a change in allocation ratio. However one needs to be aware of the effect of time trends as discussed in \cite{GreenstreetPeter2021Ammp, RoigMartaBofill2023Oasi} and one may wish to use a modelling approach \citep{lee2020including}. 
\par 
Furthermore in this work we have looked at a ideal example where the trial has equal allocation as planned. However in reality the probability of having equal allocation is very slim depending on the treatment allocation method. Therefore we have also considered the effect of using simple random allocation. This therefore means criteria of Theorem \ref{ConTheorem2} or Theorem \ref{ConTheorem3} are  no longer met. We have investigated this for three cases. We studied the number of times out of 100,000,000 simulations that keeping the data has resulted in the treatment of interest being taken forward when this would not have been the case using only the new data. This probability is still very small (0.0006\% in the example studied) relative to the probability that discarding the old data has resulted in the treatment of interest being taken forward when this would not have been the case using all the data. This can be seen in the Supplementary Materials Section 8. 

\par
Overall this paper has highlighted the importance to consider what to do if you change control during a platform trial to one which has been found superior to the control. Therefore one should consider if they should continue the current trial or stop and start a new trial with the new control.


\begin{appendices}

\section{Formulation of the events for calculating the conditional power and overall}
\label{App:Events}
The event $E^1_{k^\star,k',j'}$ which is the event that treatment $k'$ becomes the control at it's $j'^{\text{th}}$ stage equals, 
\begin{align*}
E^1_{k^\star,k',j'}=\bigcap^{j'-1}_{i=1} (l_{i} \leq Z_{k',0,i} \leq u_{i}) \cap Z_{k',0,j'} \geq u_{j'}.
\end{align*}
The event $E^2_{k^\star,k',j'}$ which is that treatment $k^\star$ is still in the trial when treatment $k'$ becomes the control equals,
\begin{align*}
& E^2_{k^\star,k',j'}= (n_{k^\star,1}>n_{k',j'}) \cup (n_{k^\star,1}\leq n_{k',j'}) \cap \bigg{\{} \bigg{[} (n_{k^\star,{\ddot{j}_{k^\star,k',j'}}} < n_{k',j'})  \cap
\bigcap^{{\ddot{j}_{k^\star,k',j'}}}_{i=1} (l_{i} \leq Z_{k^\star,0,i} \leq u_{i})\bigg{]}
\\
& \cup \bigg{[} (n_{k^\star,{\ddot{j}_{k^\star,k',j'}}} = n_{k',j'})\bigcap^{{\ddot{j}_{k^\star,k',j'}}-1}_{i=1} (l_{i} \leq Z_{k',0,i} \leq u_{i}) 
\cap (l_{{\ddot{j}_{k^\star,k',j'}}} \leq Z_{k^\star,0,{\ddot{j}_{k^\star,k',j'}}})
\\
& \cap [( Z_{k^\star,0,{\ddot{j}_{k^\star,k',j'}}} \leq u_{\ddot{j}_{k^\star,k',j'}})
\cup 
( Z_{k^\star,k,{\ddot{j}_{k^\star,k',j'}}} \leq 0 )
]
%
%
\end{align*}
The event $E^3_{k^\star,k',j'}$ which is that none of the other $k$ treatments become the control is 
\begin{align*}
E^3_{k^\star,k',j'}&= \bigcap_{k \in (1 \hdots K)/ k^\star, k'} \Bigg{(}(n_{k,1}>n_{k',j'}) \cup (n_{k,1}\leq n_{k',j'}) \cap \bigg{\{} \bigg{[}
\bigcup^{\ddot{j}_{k,k',j'}-1}_{i=1}
\bigcap^{i-1}_{i^\star=1} (l_{i^\star} \leq Z_{k,0,i^\star} \leq u_{i^\star})
\\
& \cap (Z_{k,0,i} \leq l_{i})\bigg{]} \cup \bigg{(} \bigg{[} (n_{k,{\ddot{j}_{k,k',j'}}} < n_{k',j'}) \cap \bigcap^{{\ddot{j}_{k,k',j'}}-1}_{i=1} (l_{i} \leq Z_{k,0,i} \leq u_{i}) 
\\
& \cap (Z_{k,0,{\ddot{j}_{k,k',j'}}} \leq u_{\ddot{j}_{k,k',j'}} ) \bigg{]}   \cup  \bigg{[} (n_{k,{\ddot{j}_{k,k',j'}}} = n_{k',j'})\cap \bigcap^{{\ddot{j}_{k,k',j'}}-1}_{i=1} (l_{i} \leq Z_{k,0,i} \leq u_{i}) 
\\
& \cap [(Z_{k,0,{\ddot{j}_{k,k',j'}}} \leq u_{\ddot{j}_{k,k',j'}} ) \cup (Z_{k,k',{\ddot{j}_{k,k',j'}}} < 0)] \bigg{]} \bigg{)} \bigg{\}} \Bigg{)}.
\end{align*}
The event $E^4_{k',k^\star,j'}$ which is the event that we $\text{reject } H_{k'k}$ within the rest of the trial equals,
\begin{align*}
E^4_{k^\star,k',j'}= \bigcup^J_{i=\ddot{j}_{k^\star,k',j'}+1}  \bigcap^{i-1}_{i^\star=\ddot{j}_{k^\star,k',j'}+1} (l_{i^\star} \leq Z_{k^\star,k',i^\star} \leq u_{i^\star}) \cap  (u_{i} < Z_{k^\star,k',i}).
\end{align*} 
The event $E^{\star 4}_{k^\star,k',j'}$ which is the event that we $\text{reject } H_{k'k}$ within the rest of the trial when not retaining the information post the change in control treatment equals,
\begin{align*}
E^{\star 4}_{k^\star,k',j'}=
\bigcup^J_{i=\ddot{j}_{k^\star,k',j'}+1}  \bigcap^{i-1}_{i^\star=\ddot{j}_{k^\star,k',j'}+1} (l_{i^\star} \leq Z^\star_{k^\star,k',i^\star,j'} \leq u_{i^\star}) \cap  (u_{i} < Z^\star_{k^\star,k',i,j'}) .
\end{align*} 

\section{Proof of Theorem 1 and Theorem 2}
\label{App:ConTheoremproves}
The proof of Theorem \ref{ConTheorem1} is:
\begin{proof}

Define $\hat{Z}_{k,k',j'}$, where $\hat{Z}_{k,k',j'}$ equals $Z_{k,k',j}$ at $n_{k',j'}$, so: 
\begin{align*}
\hat{Z}_{k,k',j'}=& \frac{ \sum^{n_{k',j'}}_{i=\max(n_{k,0},n_{k',0})+1} X_{k,i} - \sum^{n_{k',j'}}_{i=\max(n_{k,0},n_{k',0})+1} X_{k',i} }{\sigma\sqrt{2(n_{k',j'}-\max(n_{k,0},n_{k',0}))}}.
\end{align*}
Therefore,
\begin{align*}
Z_{k,k',j}= \frac{ \hat{Z}_{k,k',j'} \sqrt{n_{k',j'}-\max(n_{k,0},n_{k',0})} + Z^\star_{k,k',j,j'} \sqrt{n_{k,j}-n_{k',j'})}}{\sqrt{(n_{k,j}-\max(n_{k,0},n_{k',0}))}}.
\end{align*}
The same boundaries $U$ and $L$ ,as predefined for the trial, are used so if the old data is kept one can rearrange $Z_{k,k',j}>u_j$ to be: 
\begin{align*}
Z^\star_{k,k',j,j'} >  \frac{u_j \sqrt{(n_{k,j}-\max(n_{k,0},n_{k',0}))}- \hat{Z}_{k,k',j'} \sqrt{n_{k',j'}-\max(n_{k,0},n_{k',0})}}{\sqrt{n_{k,j}-n_{k',j'}}},
\end{align*}
compared to $Z^\star_{k,k',j,j'}>u_j$ for only new data.
There is only increased chance of going above $u_j$ when keeping the historic data if:
\begin{align}
\hat{Z}_{k,k',j'} >  \frac{u_j [\sqrt{(n_{k,j}-\max(n_{k,0},n_{k',0}))}- \sqrt{n_{k,j}-n_{k',j'}}]}{\sqrt{n_{k',j'}-\max(n_{k,0},n_{k',0})}}.
\label{Eq:hatZabove}
\end{align}
For an increase chance of rejecting the null hypothesis $H_{k,k'}$ at the next stage if pre change data is kept compared to discard it one requires $\hat{Z}_{k,k',j'}$ to be positive if $u_j$ is positive.
Using Equation \eqref{Eq:hatZabove} if all treatment are added at the same point it is worth keeping the historic data if:
\begin{align}
\hat{Z}_{k,k',j'} >  \frac{u_j [\sqrt{n_{k,j}}- \sqrt{n_{k,j}-n_{k',j'}}]}{\sqrt{n_{k',j'}}} \geq 0.
\label{Eq:hatZ}
\end{align}
However $\hat{Z}_{k,k',j'}<0$ as treatment $k'$ is the new control not treatment $k^*$. 
\end{proof}
The proof of Theorem \ref{ConTheorem2} is:

\begin{proof}
Define the following: 
\begin{align*}
B_{k^\star,j}(\delta_{1,j},\delta_{2,j})=& [(\delta_{2,j}l_j+\delta_{1,j}) < Z^\star_{k^\star,k',j}<(\delta_{2,j}u_j+\delta_{1,j})]
\\
C_{k^\star,j}(\delta_{1,j},\delta_{2,j})=& [(\delta_{2,j}u_j+\delta_{1,j}) < Z^\star_{k^\star,k',j}].
\end{align*}
From Definition \ref{Def:Conpower} the conditional power equals:
\begin{equation*}
R(\delta_{1,j},\delta_{2,j})= \bigcup^J_{j=j'+1}\bigg{[}\bigcap^{j-1}_{i=j'+1}(B_{k^\star,i}(\delta_{1,i},\delta_{2,i})) \cap C_{k^\star,j}(\delta_{1,j},\delta_{2,j})  \bigg{]}.
\end{equation*}
When no data is taken $\delta_{1,j}=0$ and $\delta_{2,j}=1$.
However when old data is taken forward $\delta_{1,j}=\frac{-\hat{Z}_{k,k',j'}\sqrt{n_{k',j'}}}{\sqrt{n_{k,j}-n_{k,j'}}}$ and $\delta_{2,j}=\frac{\sqrt{n_{k,j}}}{\sqrt{n_{k,j}-n_{k,j'}}}$. Therefore when old data is retained $\delta_{1,j}\geq 0$ and  $\delta_{2,j}\geq 1$ as $\hat{Z}_{k,k',j'}<0$.
\par
Then under the assumption $u_j \geq 0$ and $l_j \geq 0$ for all $j=(j'+1) \hdots J$. For any $\epsilon_{1,j} \geq 0$ and $\epsilon_{2,j} \geq 0$ let 
\begin{equation*}
w = (Z^\star_{k^\star,k',j'+1},\hdots,Z^\star_{k^\star,k',J}) \in \bigcup^J_{j=j'+1}\bigg{[}\bigcap^{j-1}_{i=j'+1}(B_{k^\star,i}(\delta_{1,i}+\epsilon_{1,i},\delta_{2,i}+\epsilon_{2,i})) \cap C_{k^\star,j}(\delta_{1,j}+\epsilon_{1,j},\delta_{2,j}+\epsilon_{2,j})  \bigg{]},
\end{equation*}
for some $q \in \{j'+1, \hdots, J \}$ for which $Z^\star_{k^\star,k',q} \in C_{k^\star,q}(\delta_{1,q}+\epsilon_{1,q},\delta_{2,q}+\epsilon_{2,q})$ and $Z^\star_{k^\star,k',h} \in B_{k^\star,h}(\delta_{1,h}+\epsilon_{1,h},\delta_{2,h}+\epsilon_{2,h})$ for $h=j'+1,\hdots q-1$. $Z^\star_{k^\star,k',q} \in C_{k^\star,q}(\delta_{1,q}+\epsilon_{1,q},\delta_{2,q}+\epsilon_{2,q})$ implies that $Z^\star_{k^\star,k',q} \in C_{k^\star,q}(\delta_{1,q},\delta_{2,q})$. Furthermore $Z^\star_{k^\star,k',q} \in B_{k^\star,q}(\delta_{1,q}+\epsilon_{1,q},\delta_{2,q}+\epsilon_{2,q})$  implies that $Z^\star_{k^\star,k',q} \in B_{k^\star,q}(\delta_{1,q},\delta_{2,q}) \cup C_{k^\star,q}(\delta_{1,q},\delta_{2,q})$ for some $h=j'+1,\hdots q-1$. Therefore, 
\begin{equation*}
w = (Z^\star_{k^\star,k',j'+1},\hdots,Z^\star_{k^\star,k',J}) \in \bigcup^J_{j=j'+1}\bigg{[}\bigcap^{j-1}_{i=j'+1}(B_{k^\star,i}(\delta_{1,i},\delta_{2,i})) \cap C_{k^\star,j}(\delta_{1,j},\delta_{2,j})  \bigg{]}.
\end{equation*}
As a result $P(R(0,1))\geq P(R(\frac{-\hat{Z}_{k,k',j'}\sqrt{n_{k',j'}}}{\sqrt{n_{k,j}-n_{k,j'}}},\frac{\sqrt{n_{k,j}}}{\sqrt{n_{k,j}-n_{k,j'}}}))$.
\end{proof}

\section{Proof of Theorem 4 and Theorem 5}
\label{App:OverTheoremproves}
The proof of both Theorem \ref{OverTheorem1} and Theorem \ref{OverTheorem2} is:
\begin{proof}
Let the treatment with the greatest positive treatment effect be treatment $k^\star$. Then one can write the conditional power of treatment k if no pre change data is kept as:
\begin{align*}
\frac{P(E^1_{k^\star,k',j'} \cap E^2_{k^\star,k',j'} \cap E^3_{k^\star,k',j'})P(E^{\star 4}_{k^\star,k',j'})}{P(E^1_{k^\star,k',j'} \cap E^2_{k^\star,k',j'} \cap E^3_{k^\star,k',j'})}.
\end{align*}
From Theorem \ref{ConTheorem2} when $u_j \geq 0$ and $l_j \geq 0$ for all $j=1 \hdots J$ is true we know for a given $k'$ and $j'$
\begin{align*}
P(E^1_{k^\star,k',j'} \cap E^2_{k^\star,k',j'} \cap E^3_{k^\star,k',j'} \cap E^{4}_{k^\star,k',j'})
\leq
P(E^1_{k^\star,k',j'} \cap E^2_{k^\star,k',j'} \cap E^3_{k^\star,k',j'})P(E^{\star 4}_{k^\star,k',j'}).
\end{align*}
Additionally this is known for when $u_j \geq 0$ and there are no lower boundaries  for all $j=1 \hdots J$ from Theorem \ref{ConTheorem3}. 
This is true for every  $k' \in \{ 1,\hdots, K \} / k^*$ and $j' \in 1,\hdots, J$, so
\begin{equation*}
\sum^{J}_{j^\star=1} \Xi_{k^\star,j^\star} + \sum_{k'\in \{1,\hdots,K \}/k^\star }\sum^{J}_{j'=1}\Omega_{k^\star,k',j'} \leq \sum^{J}_{j^\star=1} \Xi_{k^\star,j^\star} + \sum_{k'\in \{1,\hdots,K \}/k^\star }\sum^{J}_{j'=1}\Omega^\star_{k^\star,k',j'}.
\end{equation*}
\end{proof}

\end{appendices}

\section*{SOFTWARE}
Software in the form of R code can be found at:
\\
 https://github.com/pgreenstreet/change\_control\_platform.

\section*{Competing interests}
The authors declare no potential conflict of interests. Alun Bedding is a shareholder of Roche Products Ltd.

\section*{Acknowledgments}
This report is independent research supported by the National Institute for Health Research
(NIHR300576). The views expressed in this publication are those of the authors and not necessarily those of the NHS, the National Institute for Health Research or the Department of Health and Social Care (DHSC). TJ and PM
also received funding from UK Medical Research Council (MC UU 00002/14 and MC UU 00002/19, respectively). This paper is based on work completed while PG was part of the EPSRC funded STOR-i centre for doctoral training (EP/S022252/1). For the purpose of open access, the author has applied a Creative Commons Attribution (CC BY) licence to any Author Accepted Manuscript version arising.

\bibliography{refp1} 

\end{document}